\documentclass[12pt]{article}
\usepackage{graphicx}

\begin{document}
\title{
\begin{flushright} \begin{small}
  DTP--MSU/00-12 \\ LAPTH -- 822/2000 \\ hep-th/0012059
\end{small} \end{flushright}
\vspace{2.cm}
{\bf D-branes and vacuum periodicity}}

\author{Dmitri Gal'tsov\thanks{Email: galtsov@grg.phys.msu.su} $\,^{a,b}$ %
and Vladimir Dyadichev\thanks{Email: rkf@mail.ru} $\,^b$\\ \\  %
\it%
$^a$ Laboratoire de  Physique Th\'eorique LAPTH (CNRS), \\
\it
 B.P.110,F-74941 Annecy-le-Vieux cedex, France,\\
 \\  %
 \it $^b$ Department of Theoretical Physics,\\
 \it Moscow State University, 119899, Moscow, Russia,
}%

\date{\today}

\maketitle

\begin{abstract}
The superstring/M-theory suggests the Born-Infeld type modification of the
classical gauge field lagrangian. We discuss how this changes topological
issues related to vacuum periodicity in the $SU(2)$ theory in four spacetime
dimensions. A new feature, which is due to the breaking of  scale invariance
by  the non-Abelian Born-Infeld (NBI) action, is that the potential barrier
between the neighboring vacua is lowered to a finite height. At the top of
the barrier one finds an infinite family of sphaleron-like solutions
mediating transitions between different topological sectors. We review these
solutions for two versions of the NBI action: with the ordinary and
symmetrized trace. Then we show the existence of sphaleron excitations of
monopoles in the NBI theory with the triplet Higgs.  Soliton solutions in
the constant external Kalb-Ramond field are also discussed which correspond
to monopoles in the gauge theory on non-commutative space. A
non-perturbative monopole solution for the non-commutative $U(1)$ theory is
presented.
\end{abstract}

\newpage

\section{Introduction}
Recent development in the superstring theory \cite{gal-Po98,gal-GiKu98}
suggests that the low-energy dynamics of a $Dp$-brane moving in a flat
D-dimensional spacetime $z^M=z^M(x^\mu),\, M=0,...,D-1,\,\mu=0,\ldots p$ is
governed by the Dirac-Born-Infeld (DBI) action
\begin{equation}
S_p=\int\left(1-\sqrt{-\det(g_{\mu\nu}+F_{\mu\nu})}\right)d^{p+1}x,
\end{equation}
where
\begin{equation}
g_{\mu\nu}=\partial_\mu z^M\partial_\nu z^N \eta_{MN},
\end{equation}
is an induced metric on the brane and $F_{\mu\nu}$ is a $U(1)$ gauge field
strength. Using the gauge freedom under diffeomorphisms of the world-volume,
one can choose  coordinates   $z^M=(x^\mu,\, X^m)$, where $X^m$ are
transverse to the brane,  and rewrite  the action as
\begin{equation}
S_p=\int\left(1-\sqrt{-\det(\eta_{\mu\nu} + \partial_\mu X^m \partial_\nu
X^m+F_{\mu\nu})}\right)d^{p+1}x.
\end{equation}

A trivial solution to this action is $X^m=0,\, F_{\mu\nu}=0$, what means
that the $p$-brane is flat and there is no electromagnetic field. Because of
the symmetry $X^m\to -X^m$,  the planar solution remains true when
$F_{\mu\nu}$  does not vanish, in which case the electromagnetic field is
governed by the Born-Infeld (BI) action. Moreover, as was noticed by Gibbons
\cite{gal-Gi97}, the only regular static source-free solution of the BI
electrodynamics which falls off at spatial infinity is a trivial one.

This is no longer true in the case of $N$ coincident $Dp$-branes whose
low-energy dynamics is described by the non-Abelian generalization of the
DBI action involving the $SU(N)$ Yang-Mills (YM) field. Namely, for flat
$D3$-branes the regular sourceless finite energy configurations of the YM
field  were found to exist \cite{gal-GaKe99,gal-DyGa00}. The topological
reason for this lies in the vacuum periodicity of the $SU(2)$ gauge field in
four dimensions.  Neighboring YM vacua are separated by potential barriers
which in the case of the BI action are lowered down to a finite height due
to the breaking of the scale invariance in the BI theory. This removes the
well-known obstruction for classical glueballs
\cite{gal-De76,gal-Pa77,gal-Co77}, which can be summarized as follows. Scale
invariance of the usual quadratic Yang-Mills action implies that the YM
field stress--energy tensor is traceless: $T_\mu^\mu=0=-T_{00}+ T_{ii}$,
where $\mu=0,...,3,\; i=1,2,3$. Since the energy density is positive,
$T_{00}>0$, the sum of the principal pressures $T_{ii}$ is also everywhere
positive, {\it i.e.} the Yang--Mills matter is repulsive. Consequently,
mechanical equilibrium within the localized static YM field configuration is
impossible \cite{gal-Gi82}. In the spontaneously broken gauge theories scale
invariance is broken by scalar fields, what opens the possibility of
particle-like solutions: magnetic monopoles (in the theory with the real
triplet Higgs) and sphalerons (in the theory with the complex doublet
Higgs).

The role of the Higgs field in these two cases is somewhat different. For
monopoles the topological significance of the Higgs field is essential:
indeed, monopoles interpolate between the unbroken and broken Higgs phases.
In the case of sphalerons, the Higgs field  plays mostly a role of an
attractive agent which is able to glue the repulsive YM matter.
Historically, topological significance of the Dashen-Hasslacher-Neveu (DHN)
solution in the $SU(2)$ theory with the doublet Higgs~\cite{gal-DaHaNe74}
was first explained by Manton \cite{gal-Ma83} as a consequence of
non--triviality of the {\em third} homotopy group of the Higgs broken phase
manifold $\pi_3{(G/H)}$. This is equivalent to existence of non-contractible
loops in the space of field configurations passing through the vacuum. Then
by the minimax argument one finds that a saddle point exists on the energy
surface which is a proper place for the sphaleron. Later it became clear
that similar solutions arise in some models without Higgs, such as
Einstein-Yang-Mills \cite{gal-BaMc88} or Yang-Mills with dilaton
\cite{gal-LaMa93} (for a review and further references
see~\cite{gal-VoGa98}). The main common feature of these theories is that
the conformal invariance of the classical YM equations is broken, what
removes the "mechanical" obstruction for existence of particle-like
configurations. As far as the topological argument is concerned, it is worth
noting that $H=1$ for the DHN solution, so the same third homotopy group
argument applies to the gauge group $G$ itself, that is, it works equally in
the theories without Higgs.

Breaking of the scale invariance in the NBI theory also gives rise to
sphaleron glueballs which mediate transitions between different topological
sectors of the theory. Their mass is related to the BI field-strength
parameter which for the D-branes is $2\pi\alpha'$. We will discuss here
glueball solutions in two versions of the NBI theory: with the ordinary and
symmetrized trace. We also show that, when the triplet Higgs field is added,
the theory admits, apart from the usual magnetic monopoles, the hybrid
solutions which can be interpreted as sphaleron excitations of monopoles. At
the end we briefly discuss monopole solutions in gauge theories on
non-commutative spaces and give an explicit solution for the $U(1)$ monopole
with Higgs in the D-brane picture with the Kalb-Ramond field.

\section{NBI action with ordinary and symmetrized trace}

A precise definition of the NBI action was actively discussed during past
few years
\cite{gal-Ts97,gal-GaGoTo98,gal-Br98,gal-BrPe98,gal-Pa99,gal-Za00}, for an
earlier discussion see \cite{gal-Ha81}. An ambiguity is encoded in
specifying the trace operation over the gauge group generators. Formally a
number of possibilities can be envisaged. Starting with the determinant form
of the $U(1)$ Dirac-Born-Infeld action
\begin{equation}  \label{gal-det}
S=\frac{1}{4\pi}\int
\left\{1-\sqrt{-\det(g_{\mu\nu}+F_{\mu\nu})}\right\}d^4x,
\end{equation}
one can use the usual trace, the symmetrized  or antisymmetrized
\cite{gal-Ts97} ones, or   evaluate the determinant both with respect to
Lorentz and the gauge matrix indices \cite{gal-Pa99}. Alternatively one can
start with the 'square root' form, which is most easily derived from
(\ref{gal-det}) using the identities
\begin{eqnarray}
&\det(g_{\mu\nu}+F_{\mu\nu})=\det(g_{\mu\nu}-F_{\mu\nu})=
\det(g_{\mu\nu}+i{\tilde F}_{\mu\nu})=\nonumber&\\&
=\det(g_{\mu\nu}-i{\tilde F}_{\mu\nu})=
\left[\det(g_{\mu\nu}-F^2_{\mu\nu})(g_{\mu\nu}+{\tilde F}^2_{\mu\nu})\right]
^{1/4}&,
\end{eqnarray}
where $F^2_{{\mu\nu}}=F_{\mu\alpha}{F^{\alpha}}_\nu$ (similarly for
$\tilde{F}_{\mu\nu}$), and
\begin{eqnarray} \label{gal-id}
F_{\mu\alpha}{F^\alpha}_\nu-\tilde{F}_{\mu\alpha}\tilde{F}^\alpha{}_\nu&=&
\frac12 g_{\mu\nu} F_{\alpha\beta}F^{\alpha\beta},\nonumber\\
F_{\mu\alpha}\tilde{F}^\alpha{}_\nu&=& -\frac14
g_{\mu\nu}F_{\alpha\beta}\tilde{F}^{\alpha\beta}.
\end{eqnarray}

This gives the relation
\begin{equation}
\sqrt{-\det(g_{\mu\nu}+F_{\mu\nu})}= \sqrt{-\det(g)}\;\sqrt{1+\frac12 F^2
-\frac{1}{16}(F\tilde F)^2},
\end{equation}
with $F^2=F_{\mu\nu} F^{\mu\nu},\; F\tilde{F}=F_{\mu\nu}\tilde{F}^{\mu\nu}$.

For a non-Abelian gauge group the relations (\ref{gal-id}) are no longer
valid, so there is no direct connection between the 'determinant' and the
'square root' form of the lagrangian. Therefore the latter can  be chosen as
an independent starting point for a non-Abelian generalization.

There is, however, a particular trace operation -- symmetrized trace --
under which generators commute, so both forms of the lagrangian remain
equivalent. This definition is favored by the no-derivative argument, as was
clarified by Tseytlin \cite{gal-Ts97}. Restricting the validity of the
non-Abelian effective action  by the constant field approximation, one has
to drop commutators of the matrix-valued $F_{\mu\nu}$ since these can be
reexpressed through the derivatives of $F_{\mu\nu}$. This corresponds to the
following definition
\begin{equation}
\label{gal-Strdet} S=\frac{1}{4\pi}\ensuremath{\mathop{\rm
Str}\nolimits}\int \left\{1-\sqrt{-\det(g_{\mu\nu}+F_{\mu\nu})}\right\}d^4
x,
\end{equation}
where symmetrization applies to the field strength (not to potentials). This
action reproduces an exact string theory result for non-Abelian fields up to
$\alpha'^2$ order. Although there is no reason to believe that this will be
true in higher orders in $\alpha'$, the $\ensuremath{\mathop{\rm
Str}\nolimits}$ action is an interesting model providing minimal
generalization of the Abelian action \cite{gal-Ts97}.

An explicit form of the  SU(2) NBI action with the symmetrized trace for
static $SO(3)$-symmetric magnetic type configurations was found only
recently \cite{gal-DyGa00}. One starts with the definition
\begin{equation} \label{gal-LNBI}
L_{NBI}=\frac{\beta^2}{4\pi} \ensuremath{\mathop{\rm Str}\nolimits}\left(1-
\sqrt{-\det\Bigl(g_{\mu\nu}+ \frac{1}{\beta}F_{\mu\nu}\Bigr)}\right)= k
\frac{\beta^2}{4\pi}\ensuremath{\mathop{\rm Str}\nolimits} (1-\mathcal{R}),
\end{equation}
where
\begin{equation}
\mathcal{R}=\sqrt{1+\frac{1}{2\beta^2}F_{\mu\nu}F^{\mu\nu}
-\frac{1}{16\beta^4}(F_{\mu\nu}{\tilde F}_{\mu\nu})^2},
\end{equation}
and $\beta$ of the dimension of length$^{-2}$ is the BI 'critical field'.
The normalization of the gauge group generators is unusual and is chosen as
follows
\begin{equation}
F_{\mu\nu}=F^a_{\mu\nu} t_a,\quad \ensuremath{\mathop{\rm tr}\nolimits} t_a
t_b =\delta_{ab}.
\end{equation}
The symmetrized trace of the product of $p$ matrices is defined as
\begin{equation}
\ensuremath{\mathop{\rm Str}\nolimits}(t_{a_1}\dots
t_{a_p})\equiv\frac{1}{p!} \ensuremath{\mathop{\rm
tr}\nolimits}\left(t_{a_1}\dots t_{a_p} + \mbox{all permutations}\right),
\end{equation}
and it is understood that the general matrix function like (\ref{gal-LNBI})
has to be series expanded. It has to be noted that   under the
$\ensuremath{\mathop{\rm Str}\nolimits}$ operation the generators can be
treated as commuting objects, and the gauge algebra should not be applied,
(e.g. the square of the Pauli matrix $\tau_x^2\neq 1$) until the
symmetrization in the series expansion is completed.

A general  $SO(3)$ symmetric $SU(2)$ gauge field  is described by the
Witten's ansatz
\begin{equation}\label{gal-wittans}
\sqrt{2}A = a_{0}t_{1}\; dt + a_{1}t_{1}\; dr +\{w_{2}\; t_{2} - (1-w)\;
t_{3}\}\ d\theta +\{(1-w)\ t_{2} + \tilde w \; t_{3}\}\sin\theta\; d\phi,
\end{equation}
where the functions $a_0,\,a_1,\, w,\, \tilde w$ depend on $r,t$ and
$\sqrt{2}$ is introduced to maintain the standard normalization. Here we use
a rotating basis $t_i,\, i=1,2,3$ for the $SU(2)$ generators defined as
\begin{equation}
t_1=n^{a}\tau^{a}/\sqrt{2},\quad t_{2}=\partial_{\theta}t_{1},\quad
\sin\theta t_{3}=\partial_{\varphi}t_{1},
\end{equation}
where $n^{a}=(\sin\theta  \cos\varphi,\sin\theta\sin\varphi, \cos\theta)$,
with $\tau^a$ being the Pauli matrices. These generators obey the
commutation relations $[t_i,t_j]=\frac{1}{\sqrt{2}}\epsilon_{ijk}t_k$.

From four functions entering this ansatz one can be gauged away. In the
static case we can further reduce the number of independent functions to
two, while the static purely magnetic configurations are fully described by
a single function  $w(r)$:
\begin{equation}\label{gal-Acompts}
\sqrt{2}  A_{\theta}=-(1-w)t_3,\qquad
\sqrt{2}A_{\varphi}=\sin\theta(1-w)t_2. \qquad A_t=A_r=0.
\end{equation}

The field strength tensor has the following non-zero components
\begin{equation} \label{gal-Fcompts}
 \sqrt{2} F_{r \theta}=w't_3,\quad  \sqrt{2}F_{r \varphi}=-\sin\theta w' t_2,
  \quad \sqrt{2}F_{\theta \varphi}=\sin\theta(w^2-1)t_1,
\end{equation}
where prime denotes derivatives with respect to $r$.

For purely magnetic configurations the second term under the square root is
zero, and the substitution of (\ref{gal-Fcompts}) gives
\begin{equation}
\mathcal{R}^2=1+\frac{(1-w^2)^2}{\beta^2 r^4} t_1^2 +\frac{w'^2}{\beta^2
r^2} (t_2^2+t_3^2).
\end{equation}
To find an explicit expression for the lagrangian one has to expand the
square root in a triple series in terms of the even powers of generators
$t_1, t_2, t_3$, then to calculate the symmetrized trace of the powers of
generators in all orders, and finally to make a resummation of the series.
The result reads
\begin{equation} \label{gal-Str}
L_{NBI}=\frac{\beta^2}{4\pi}\left(1-\frac{1+V^2+K^2\mathcal{A}}{\sqrt{1+V^2}}
\right),
\end{equation}
where
\begin{eqnarray}\label{gal-cAdef}
V^2&=&\frac{(1-w^2(r))^2}{2\beta^2r^4},\quad K^2=\frac{w'^2(r)}{2\beta^2
r^2},\nonumber\\ \mathcal{A}&=&\sqrt{\frac{1+V^2}{V^2-K^2}}
\ensuremath{\mathop{\mathrm{arctanh}}}\sqrt{\frac{V^2-K^2}{1+V^2}}.
\end{eqnarray}
Here we assumed that $V^2>K^2$, otherwise an $\arctan$ form is more
appropriate. Note that when the difference $V^2=K^2$ changes sign, the k
function $\mathcal{A}$ remains real valued. It can be checked that when
$\beta \to \infty$, the standard Yang-Mills lagrangian (restricted to
monopole ansatz) is recovered. In the strong field region our expression
differs essentially from the square root/ordinary trace lagrangian.

The corresponding explicit action defined in a square root form with an
ordinary trace reads:
\begin{equation}
L_{NBI}=\frac{\beta^2}{4\pi}\left(1-\sqrt{1+V^2+2K^2}\right)\label{gal-Otr}
\end{equation}

\section{Topological vacua and sphalerons}

As is well-known, vacuum in the $SU(2)$ YM theory in the four-dimensional
spacetime splits into an infinite number of disjoint classes which can not
be deformed into each other by 'small' (contractible to a point) gauge
transformations. Writing the pure gauge vacuum YM potentials as
$A=i\mathrm{U}d\mathrm{U}^{-1}$, where $\mathrm{U}\in SU(2)$ and imposing an
asymptotic condition
 \begin{equation} \lim_{r\to \infty} \mathrm{U}(x^i)=\mathbf{1} ,
\label{gal-Uinfty}
\end{equation}  we can interpret
$\mathrm{U}(x^i)$ as mappings $S_3  \rightarrow SU(2)$. All sets of such
$\mathrm{U}$'s falls into the  sequence of homotopy classes characterized by
the winding number
\begin{equation}
  k[\mathrm{U}]=\frac{1}{24\pi^2}\ensuremath{\mathop{\rm tr}\nolimits}
  \int\limits_{R_3}\;\mathrm{U}d\mathrm{U}^{-1}\wedge
  \mathrm{U}d\mathrm{U}^{-1}\wedge
  \mathrm{U}d\mathrm{U}^{-1}.
 \end{equation}

A representative of the $k$-th  class can be chosen as
 \begin{equation}\mathrm{U}_k = \exp \{ i\alpha(r)t_1/\sqrt{2}
\},\quad\mbox{where} \quad\alpha(0) = 0, \alpha(\infty) =- 2\pi k.
\end{equation}
The corresponding potential will be given by the Witten ansatz with $a=0,
w=exp(i\alpha(r))$. The  asymptotic condition (\ref{gal-Uinfty}) leads to
the following fall-off  requirements.
\begin{equation} A_a = o(r^{-1} ) \quad \mbox{for}\quad  r \to  \infty
\label{gal-Ainfty}.
\end{equation}

The representatives of different vacuum classes with different $k$ cannot be
continuously deformed into each other within the class of the purely vacuum
fields. But there exists an interpolating sequence of nonvacuum field
configurations of  finite energy (the latter can be defined on shell and
then continued off-shell) satisfying the required boundary conditions
(\ref{gal-Ainfty}) that connects  different vacuum classes.  Finite energy
solutions for the actions (\ref{gal-Str}) or (\ref{gal-Otr}) should satisfy
the following boundary conditions near the origin
\begin{equation} \label{gal-zero}
w=1+b\,r^2+O(r^4),
\end{equation}
and at the infinity
\begin{equation} \label{gal-infty}
 w=\pm 1+\frac{c}{r}+O(\frac{1}{r^2}),
\end{equation}
where $b$ and $c$ are free parameters. (The value $w(\infty)=0$ together
with  finiteness of the energy implies that $w\equiv 0$.) The leading terms
are the same as required for the vacuum configurations. These solutions, if
exists, can be shown to lie on the path in the solution space connecting two
topologically distinct vacua.  Consider a one-parameter sequence of field
configurations (off shell generally) depending on a continuous parameter
$\lambda\in [0,\;\pi]$ \cite{gal-GaVo91}
\begin{equation}
A[\lambda]=i\frac{1-w}{2}\mathrm{U}_+d\mathrm{U}_+^{-1}+
i\frac{1+w}{2}\mathrm{U}_-d\mathrm{U}_-^{-1}, \label{gal-Aseq}
\end{equation}
where
\begin{equation} \mathrm{U}_{\pm}
=\exp\left\{i\lambda(w\pm 1){t}_1/\sqrt{2}\right\}.
\end{equation}
This field  vanishes for $\lambda=0$, whereas for $\lambda=\pi$ it can be
represented as
\begin{equation}                                                  \label{gal-5.8}
A[\pi]=i{\rm U}d{\rm U}^{-1},\ \ \ {\rm with}\ \ \ {\rm
U}=\exp\{i\pi(w-1)t_1/\sqrt{2}\}.
\end{equation}
In view of the above boundary condition for $w$, in the case $w(\infty)=-1$
one has the $k=1$ vacuum. Now, the crucial thing is that for $\lambda
=\pi/2$ we come back to the configuration (\ref{gal-Acompts}). So if the
solution to  the classical field equations with the required asymptotics
exists indeed, this can be interpreted as a manifestation of the finiteness
of the potential barrier between distinct vacua.

Note that the same reasoning holds for the ordinary Yang--Mills system. But
due to the scale invariance of this theory there is no function $w$ which
minimizes the energy functional.

Both the analysis of the equations following from NBI lagrangians
(\ref{gal-Str},\ref{gal-Otr}) using the methods of dynamical systems
{\cite{gal-GaKe99}} and numerical experiments \cite{gal-DyGa00} shows that
such solutions exist in both NBI models --- with ordinary and symmetrized
trace. They form a discrete sequence labeled by the number of nodes of the
function $w(r)$, and the lower one-node solution is similar to the sphaleron
of the Weinberg-Salam theory.

In the NBI theory $\beta$ is the only dimensionful parameter giving  a
natural scale of length, i.e. theories with different values of $\beta$ are
equivalent up to rescaling. Setting $\beta=1$ we obtain the equations of
motion for the symmetrized trace NBI model
\begin{equation}
\label{gal-Streqw}
\frac{d}{dr}\left\{\frac{w'}{2(V^2-K^2)}\left(\frac{K^2\sqrt{1+V^2}}{1+K^2}-
\frac{(2V^2-K^2)\mathcal{A}}{\sqrt{V^2-K^2}}\right)\right\} =
\frac{w\,V(K^2\mathcal{A} -V^2)}{(V^2-K^2)\sqrt{1+V^2}}.
\end{equation}

For the ordinary trace model one has
 \begin{equation}
 \label{gal-treqw}
 \frac{d}{d r}\left\{\frac {w'}{\sqrt {1+V^2+2K^2}}\right\}=
 -\frac {w\,V}{\sqrt {1+V^2+2K^2}},
 \end{equation}

We are looking for the solutions satisfying the boundary conditions
(\ref{gal-zero},\ref{gal-infty}). For large $r$ both equations reduce to
that of the usual YM theory, so the solutions are not much different in the
far zone. Near the origin the equations are different, more careful analysis
reveals that the nature of stationary points associated with the origin is
different for two versions of the theory.

A trivial solution to these equations (valid for both models) is an embedded
abelian monopole $w=0$. In the BI theory it has the finite energy. From the
general analysis, as discussed in \cite{gal-VoGa98} for the ordinary trace,
one finds that $w$ can not have local minima for $0<w<1, \, w<-1$ and can
not have local maxima for $-1<w<0,\,w>1$. The same remains true for the
symmetrized trace. Thus any solution which starts at the origin on the
interval $-1<w<1$ must remain within the strip $-1<w<1$. Once $w$ leaves the
strip, it diverges in a finite distance. Regular solutions exist for a
discrete sequence of $b$ shown in the table \ref{gal-tab:param}  together
with corresponding masses $M_n$ for the first  six $n$ which is the number
of zeroes of $w(r)$. The $n=1$ solution is analogous to the sphaleron known
in the Weinberg-Salam theory \cite{gal-DaHaNe74,gal-Ma83}, it is expected to
have one decay mode. Higher odd-$n$ solutions may be interpreted as excited
sphalerons, they are expected to have $n$ decay directions. Even-$n$
solutions are topologically trivial, they can be regarded as sphaleronic
excitation of the vacuum. Qualitatively picture is the same as for the
ordinary trace \cite{gal-GaKe99}, but the discrete values of $b$ are rather
different.

\begin{figure}[t]\centering
\includegraphics[width=10cm]{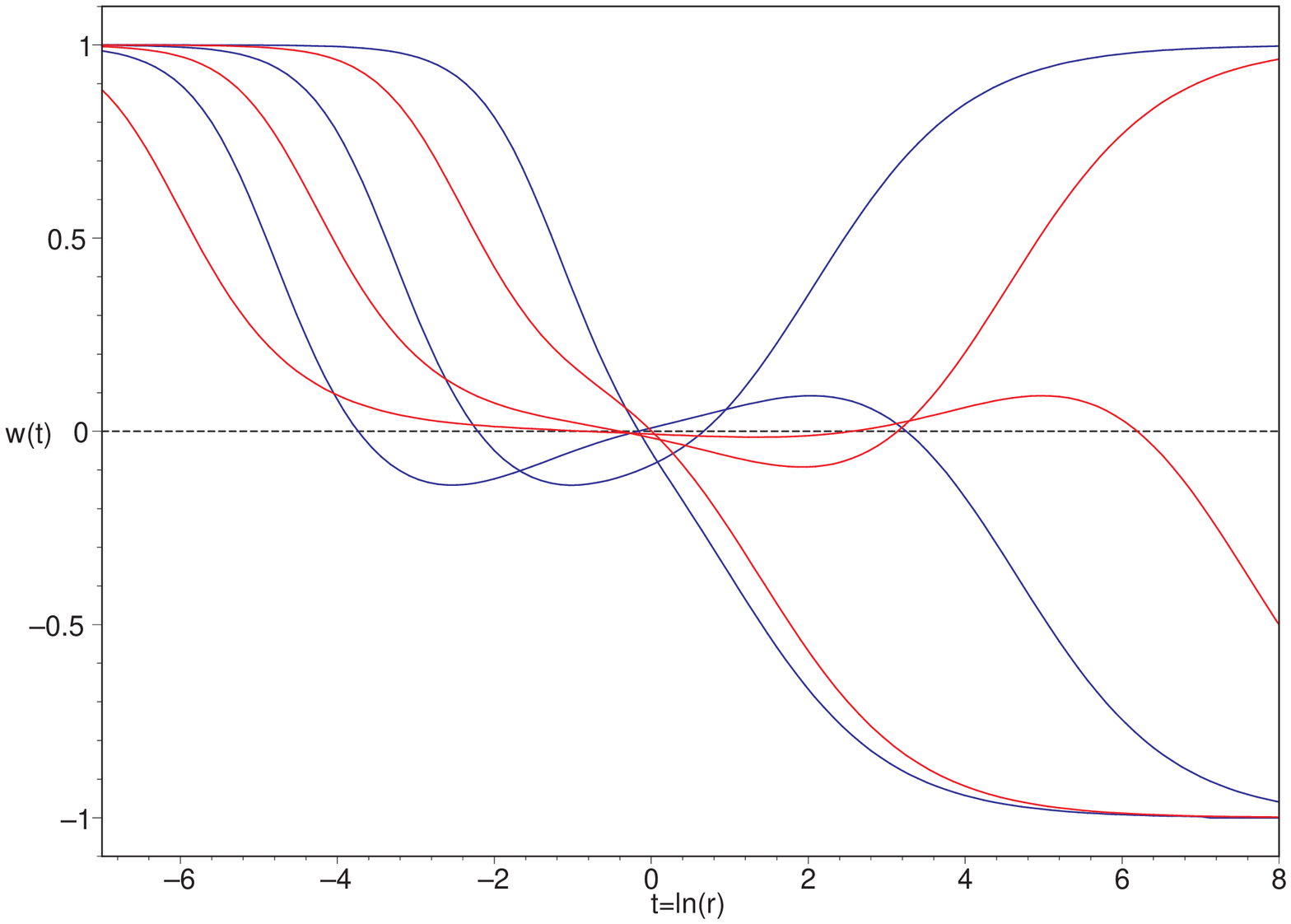}
\caption{Sphaleron glueball solutions $w_n$ for $n=1,2,3$ in the symmetrized
trace (red) and ordinary trace (blue) models} \label{gal-fig:sphalsols}
\end{figure}

Numerical solutions for both models are shown in the figure
\ref{gal-fig:sphalsols}. It is surprising that the solutions with the
ordinary and the symmetrized trace are rather similar in spite of the
substantial difference of the lagrangians. They have however somewhat
different behavior near the origin: those with the symmetrized trace leave
the vacuum value $w=1$ faster and stay  longer in the intermediate region
where $w(r)$ is close to zero. In this region the  magnetic charge is almost
unscreened, so this is the particle core. Thus for all $n$ solutions are
more compact in the ordinary trace case. For both models the parameters
$b_n$ grow infinitely with increasing node number $n$. This means that there
is no limiting solution as $n\to \infty$ contrary to the EYM case where such
solutions do exist.

\begin{table}\begin{center}
\begin{tabular}{l l l l l}
\hline  & \multicolumn{2}{c}{Ordinary trace}& \multicolumn{2}{c}{Symmetrized
trace}\\  \hline
  \multicolumn{1}{c}{$n$} &\multicolumn{1}{c}{$\quad b_{tr}$ }&
  \multicolumn{1}{c}{ $\quad M_{tr}$}&\multicolumn{1}{c}{$\quad b_{Str}$}&
  \multicolumn{1}{c}{ $\quad M_{Str}$}
\\\hline $1$ &$\quad 1.27463 \times 10^1$& $\quad 1.13559$& $\quad 1.23736\times 10^2$    & $\quad 1.20240$
\\ $2$ &$\quad 8.87397 \times 10^2$& $\quad 1.21424$& $\quad 5.05665\times 10^3$    & $\quad 1.234583$
\\ $3$ &$\quad 1.87079 \times 10^4$& $\quad 1.23281$& $\quad 1.67739\times 10^5$    & $\quad 1.235979$
\\ $4$ &$\quad 1.27455 \times 10^6$& $\quad 1.23572$& $\quad 7.11885\times 10^6$    & $\quad 1.236046$
\\ $5$ &$\quad 2.65030 \times 10^7$& $\quad 1.23603$& $\quad 4.94499\times 10^8$    & $\quad 1.2360497$
\\ $6$ &$\quad 1.80475 \times 10^9$& $\quad 1.23604$& $\quad 4.52769\times 10^{10}$ & $\quad 1.2360497$
\\\hline
\end{tabular}
\end{center}
\caption{Values of $b$ and $M$ for first six glueball solutions in NBI
models with ordinary and symmetrized traces}\label{gal-tab:param}
\end{table}

\section{Magnetic monopoles and hybrid solutions}

Magnetic monopoles are associated with  the deformed D3-branes with non zero
transverse coordinates $X^m$ interpreted as Higgs scalars. The deformation
can be thought of as caused by an open string attached to the brane. In the
BPS limit the solutions are the same as for the quadratic YM theory
\cite{gal-Br98,gal-BrPe98} Monopoles for the ordinary trace model were
constructed by Grandi, Moreno and Schaposnik \cite{gal-GrMoSc99}. For
monopoles the function $w$ monotoneously varies from the value $w=1$ at the
origin to the asymptotic value $w=0$ at infinity. Note, that assuming the
asymptotic value $w=0$ for pure gauge NBI theory we will get only embedded
abelian solution $w\equiv 0$. Our aim here is to show that, in addition,
there are hybrid NBI-Higgs solutions for which the function $w(r)$
oscillates in the core region. In other words, starting from the vacuum
$w=1$ at the origin the function $w(r)$ tries to follow the sphaleronic
behavior, but finally turns back to the monopole regime.

Adding to the NBI action the Higgs term $S=S_{NBI}+S_H$ where $S_H$ is taken
in the usual form
\begin{equation}
S_H= \frac{1}{8\pi}\int\left(D_\mu\phi^a D^\mu\phi^a-\frac{\lambda}{2}
\left(\phi^a\phi^a-v^2\right)\right),
\end{equation}
one obtains the NBI-Higgs theory, containing,  apart from $\beta$, the
second parameter $\lambda$ (without loss of generality we put the gauge
coupling constant equal to unity). For spherically symmetric static purely
magnetic configurations the YM ansatz remains the same , while for the Higgs
field
\begin{equation}
 \phi^a=\frac{H(r)}{r}n^a.
\end{equation}

For simplicity we consider here the square root form of the NBI action
(\ref{gal-Otr}). Performing an integration over spherical angles one obtains
the energy functional (equal to minus action for static configurations)
\begin{equation}
E = 4\pi \int dr\, r^2 \left\{ 2\beta^2({\cal R} - 1) +
 \frac{1}{2 r^2} \left( (H'-\frac{H}{r})^2 +
\frac{2}{r^2} H^2w^2 \right) +
 V\right\},
 \label{gal-energy}
\end{equation}
 where
\begin{equation}
{\cal
 R}=\sqrt{1+\frac{1}{\beta^2 r^{4}}(r^2 w'^{2}+
\frac{1}{2}(w^{2}-1)^{2})},\quad V= \frac{\lambda}{4}\left( \frac{H^2}{r^2}
- 1 \right)^2.
\end{equation}

Varying this functional one  finds the   equations of motion
\begin{eqnarray}
& & r^{2}w'' = \ensuremath{\mathit{w}}(\mathcal{R} H^{2} +
\ensuremath{\mathit{w}}^{2}-1)+
r^{2}\frac{\mathcal{R}'}{\mathcal{R}}\ensuremath{\mathit{w}}',
\label{gal-eqf}\\ & & r^2 H'' =2H\ensuremath{\mathit{w}}^{2}-\lambda
H(r^{2}-H^{2})\;. \label{gal-eqh}
\end{eqnarray}

Boundary conditions at infinity for a solution with  a unit magnetic charge
read
\begin{equation}
\lim_{r \to \infty} \ensuremath{\mathit{w}}(r) = 0, \qquad  \lim_{r \to
\infty} \frac{H(r)}{r}  =  1, \label{gal-boundinf}
\end{equation}
while at the origin
\begin{equation}
\ensuremath{\mathit{w}}(0) = 1, \qquad H(0) = 0 . \label{gal-orgbound}
\end{equation}

Starting with (\ref{gal-orgbound}) one can construct the following power
series solution converging in a non-zero domain around the origin:
\begin{eqnarray}
\ensuremath{\mathit{w}} =1-b r^2+ \frac{\beta b^2 \left(22 b^2 +
\beta^2\right)+ d ^2 \left(6 b^2+ \beta^2 \right)^{\frac{3}{2}}} {10\beta
\left(2b^2+ \beta^2 \right)}\,r^4 + O(r^6) \label{gal-fexp}\\ H
=d\,r^2-\left( \frac{1}{10} \lambda d+\frac{2}{5}d\, b \right )r^4 +
O(r^6),\label{gal-hexp}
\end{eqnarray}
where $b$ and $d$ are free parameters. For $\beta \to \infty$ the theory
reduces to the standard YMH-theory, admitting monopoles. In
\cite{gal-GrMoSc99} it was shown that monopole  solutions to the
Eqs.(\ref{gal-eqf}, \ref{gal-eqh}) continue to exist up to some limiting
value $\beta_{cr}$.

Now we have to explain why one can expect to have also the hybrid solutions.
Near the origin the Higgs field is close to zero, so the influence of the
term $H^2K\mathcal{R}$ is negligible, and the YM field behaves like in the
pure NBI case.  As was argued in \cite{gal-GaKe99}, NBI theories with
different $\beta$ are equivalent up to rescaling, and so for $\beta$ large
enough the solution starts forming just near the origin. But for larger $r$
the role of Higgs is increased, so one can expect that some solutions can be
trapped to the monopole asymptotic regime.  More precisely, in the region of
$r\approx 1/\sqrt{\beta}$, the function $\ensuremath{\mathit{w}}(r)$ is
similar to the sphaleron  solution of \cite{gal-GaKe99}: starting  with
$\ensuremath{\mathit{w}}=1$ it passes through $\ensuremath{\mathit{w}}=0$
and then tends to the value $\ensuremath{\mathit{w}}=1$. After leaving this
region the solution enters the region where it has properties of the NBI
monopole and at $r\to \infty$ both field functions tend to their
asymptotical values (\ref{gal-boundinf}). The Higgs field $H(r)$ for these
hybrid solutions behaves qualitatively in the same way as for the monopoles.

To obtain hybrid solutions numerically we  introduce the logarithmic
variable $t=\ln(r)$ and apply a shooting strategy to find the values of
parameters $b$ and $d$ ensuring  the monopole asymptotic conditions
(\ref{gal-fexp}-\ref{gal-hexp}) after several oscillations of
$\ensuremath{\mathit{w}}$. As an initial guess for $b$ one can take the
(appropriately rescaled for given $\beta$) glueball values found in
\cite{gal-GaKe99}. Another parameter $d$ turns out to be weakly sensitive on
$\beta$ for $\beta$ large enough. The resulting solutions for $n=1,2$ and
$\lambda=1/2$ are shown on Fig.~1,2 together with the ground state monopole
($n=0$). The masses increase with $n$ and converge rapidly to the mass of an
embedded Abelian solution with frozen Higgs:
\begin{equation}
\ensuremath{\mathit{w}}(r)\equiv 0, \qquad H(r)\equiv r \label{gal-absol}.
\end{equation}

Although this singular solution  does not satisfy the boundary conditions
(\ref{gal-orgbound}) it has finite energy within the NBI-Higgs theory, which
can be obtained by substituting the Eq. (\ref{gal-absol}) into the Eq.
(\ref{gal-energy}):
\begin{equation}
 E_{lim} =2\int \beta^2 (\mathcal{R}-1)r^2 dr=\sqrt{\beta} \int
\left(\sqrt{4+\frac{2}{x^4}}-2\right)x^2 dx=
1.467338\sqrt{\beta}.\label{gal-abmass}
\end{equation}

\begin{figure}[t]\centering
\includegraphics[width=10cm]{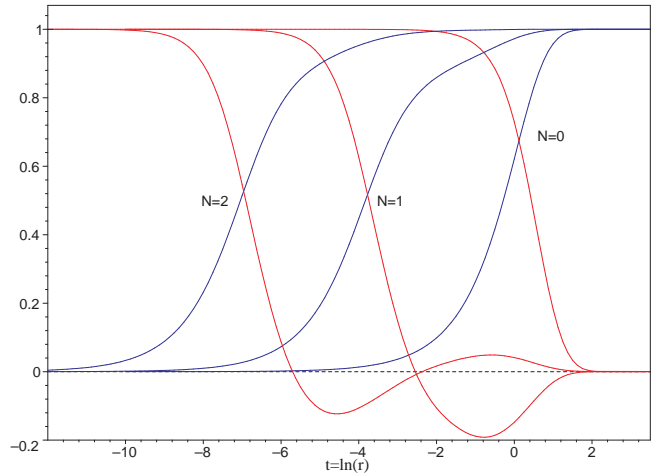}
\caption{Magnetic monopole and first two hybrid solutions in the ordinary
trace model for $\beta=30$, $\lambda=1/2$. Red line --- $w$, blue line ---
$H/r$} \label{gal-fig:exmon}
\end{figure}

With decreasing $\beta$, the discrete values of the parameter $b_n$ also
decrease until relatively small values of $\beta$. Then, with $\beta$
further decreasing both parameters $b$ and $d$   start growing until some
critical value of $\beta_{cr\,n}$ is reached near which parameters $b_n$ and
$d_n$ tend to infinity and monopole solutions with given number of zeroes
cease to exist. The lowest of these critical values is $\beta_{cr\,0}\approx
0.45$ for unexcited monopole solution. The excited solutions disappear at
greater values of $\beta$.  The mass of excited monopoles is well described
by the formula (\ref{gal-abmass}), even for the lowest excited solution the
difference with the exact numerical value is less then 4\% for all values of
$\beta$. The figure \ref{gal-fig:exmon} shows the behavior of functions
$w,\,H$   for some intermediate value of $\beta$. Note, that at critical
$\beta$ all branches of monopole solutions (including unexcited branches)
converge to the limiting Abelian solution (\ref{gal-absol}) (with different
rate).

The excited monopole solutions also exist in the Einstein-Yang-Mills-Higgs
theory \cite{gal-BrFoMa95}. There the role of non-linear excitations is
played by Bartnick-McKinnon gravitating sphalerons of EYM theory
\cite{gal-BaMc88}. The phase diagram (regions of existence in parameter
space) is somewhat different in our case, the details will be given
elsewhere.

\section{Non-commutative monopoles}
Here we discuss one another aspect of the D-brane picture of gauge theories,
which is the direct subject of the present workshop. Recently it was
discovered that gauge theories on noncommutative manifolds
\begin{equation}
[x_\mu,x_\nu]=i\theta_{\mu\nu}
\end{equation}
are connected with the gauge theories on D-branes with  the constant
background Kalb-Ramond field $B$ turned on \cite{gal-CoDoSc98}
\begin{equation}
B_{\mu\nu}=-\frac{\theta_{\mu\nu}}{(2\pi\alpha')^2}.
\end{equation}
The relation between these two versions  is non-local and is defined
perturbatively through the Seiberg-Witten map \cite{gal-SeWi99} (for a more
recent discussion see  \cite{gal-Te00,gal-Ne00}). Namely, the YM theory on a
noncommutative four-dimensional space
\begin{equation}
\hat{S}={\rm Tr} \int\left(\frac1{4{\hat g}^2}{\hat F}_{\mu\nu}\;*\; {\hat
F}^{\mu\nu}+...\right)d^4x,
\end{equation}
defined using the star-product
\begin{equation}
F(x)\; *\;G(x)=\exp\left(\frac{i\theta_{\mu\nu}}{2}
\partial_\mu\partial'_\nu\right)F(x)G(x')|_{x'=x},
\end{equation}
and the D-brane theory with $A_\mu,\,F_{\mu\nu}$ are related perturbatively
via
\begin{equation}
{\hat A}_\mu=A_\mu-\frac{\theta^{\alpha\beta}}{4}\left\{ A_\alpha,\,
\partial_\beta A_\mu+F_{\beta\mu}\right\}_+ +O(\theta^2).
\end{equation}

The issue of magnetic monopoles in both treatments of the
non-commu\-ta\-tive YM was discussed recently in a number of papers
\cite{gal-HaHa99,gal-GoHa00,gal-HaHaMo99,gal-Mo00,gal-GoHa00}. It was argued
that BPS-saturated mono\-poles exist in the non-commutative case as well.
Apart from the BPS bound most of the previous discussion was perturbative in
terms of the non-commutativity parameter $\theta_{\mu\nu}$.

Adding the constant $B$-field spoils the spherical symmetry of mono\-poles
and therefore their non-perturbative treatment in the D-brane picture
becomes rather complicated. At best one can construct an axially symmetric
model using $B_{\mu\nu}$ as a Kalb-Ramond analog of the homogeneous magnetic
(electric) field. Even in this case the NBI model is still too complicated
both for Tr and Str versions. Here we give a non-perturbative monopole
solution in the simplest case of the $U(1)$ gauge field with Abelian Higgs.
As was shown by Gibbons \cite{gal-Gi97}, the system of BI $U(1)$ and Higgs
fields possesses the boost symmetry (in the mixed space of coordinates and
the field variables) which can be used as a solution generating technique to
add a constant magnetic field to the pointlike magnetic monopole (resp.
electric field to the electric BIon). Reinterpreted as the Kalb-Ramond
field, this homogeneous field may be accounted for the parameter of
non-commutativity.

We start with the DBI action
 \begin{equation}
   S_{DBI}=-\int\; d^4x\sqrt{ -\det \left(\eta_{\mu\nu}+\partial_\mu y \partial_\nu y +
   F_{\mu\nu}\right)}
 \end{equation}
with one external coordinate $y$ (playing the role of the Higgs field) and
introduce the magnetic potential $\chi$
 \begin{equation}
 \mathbf{H}=-\nabla \chi, \end{equation}
where $\mathbf{H}$ is the magnetic field strength --- canonical conjugate to
the magnetic induction $\mathbf{B}$:
 \begin{equation}
 \mathbf{H}=-\frac{\partial L}{\partial\mathbf{B}}.
 \end{equation}

Performing the corresponding Legendre transformation we obtain the following
hamiltonian functional
 \begin{equation}
{\cal H}= \int\;d^3x\sqrt{1-(\nabla\chi)^2+(\nabla
y)^2+(\nabla\chi)^2(\nabla y)^2-(\nabla \chi\cdot\nabla y)^2},
 \end{equation}
which can be interpreted as the volume of the three-dimensional hypersurface
parametrized by coordinates $x^i$ in the five-dimensional pseudoeuclidean
space $\{x^i,y,\chi\}$ with the metric $\mathrm{diag}(+,+,+,+,-)$ (minus
corresponds to $\chi$). We use the symmetries of this functional to generate
first the scalar charge from the monopole charge and then to generate a
constant background field which will be then interpreted as the $B$ field.
So we start with the spherically symmetric configurations. The field
equations are then reduced to
\begin{equation}
y''=2\,\frac {y'\left (\chi'^2-y'^2 -1\right)}{r},\quad \chi''=2\,\frac
{\chi'\left (\chi'^2-y'^2-1\right)}{r},
\end{equation}
where prime denotes the derivative with respect to the radial variable $r$.
It is easy to see that two potentials should be proportional. Depending on
which potential dominates, one can find three different types of behaviour:
\begin{enumerate}
\item The spacelike vector in the $\{y, \chi\}$ plane. By some
rotation the magnetic field can be removed. This is the catenoidal solution
\cite{gal-Gi97}. Since it does not exist for all $r$, we will not consider
it further.

\item The timelike vector in the  $\{y, \chi \}$ plane. By
a rotation it can be reduced to a $U(1)$ monopole without excitations of the
transverse degrees of freedom. The  potential for this particular solution
(with unit charge) is
\begin{equation} \chi_0(r)=  \int\limits^{\,r}
\frac{dr}{\sqrt{1+r^4}},
 \label{gal-deff}
\end{equation}
and could be written explicitly in terms of elliptic integrals.

\item The lightlike vector $y=\pm \chi$. This is the BPS monopole:
\begin{equation}
  \chi_{BPS}(r)=\pm y_{BPS}(r)=\frac{1}{r}.
\end{equation}
\end{enumerate}

To obtain the non-BPS monopole solution that also has a nonzero Higgs
counterpart $y(r)$ one can simply perform a boost in the $\{\chi,y\}$ plane:
 \begin{equation}
   \chi(r)=\cosh \psi \,\chi_0(r) \qquad y(r)= \sinh \psi \, \chi_0(r).
 \end{equation}
The next step is to perform a boost in the $\{\chi,z\}$ plane to generate
the constant background magnetic field. To understand why this  field may be
equally interpreted as a $B$ field one should notice that the field
equations do not change if we replace $F_{\mu\nu}$ by
$F_{\mu\nu}+B_{\mu\nu}$ with constant $B$.

So, if we denote $\chi=g(\rho,z)$, then after the second boost we obtain:
 \begin{equation}
\cosh\phi\,g+\sinh\phi\,z=\cosh\psi \,
\chi_0\left(\sqrt{\rho^2+\left(\cosh\phi\,z+\sinh\phi\,g\right)^2}\right),
\label{gal-boost:eq}
 \end{equation}
where $\rho=\sqrt{x^2+y^2}$ and $\chi_0$ is defined by the
Eq.(\ref{gal-deff}).

This nonlinear equation cannot be solved explicitly but it is simple to
explore it numerically. The key point is to note that for a given
$g,\rho,z$, using equations (\ref{gal-deff}),(\ref{gal-boost:eq}), one can
find the vector $\mathbf{F}+\mathbf{B}$ (magnetic {\em induction} plus
$B$-field). Then the monopole field is obtained by subtracting the constant
background. Note that, depending on the values of the boosts parameters
$\phi$ and $\psi$, the solution can become double-valued. Let us consider
this feature in more detail. For magnetic monopole without excitations of
the transversal component the three-dimensional hypersurface $\chi_0(x,y,z)$
is spacelike everywhere except for the origin where it touches the
lightcone. When we boost in the $\{\chi,y\}$ directions, the surface
$\chi(x,y,z)$ acquires the timelike piece which can cause multivaluedness
after boosting in the $\{\chi,z\}$ directions. (When treated as a
hypersurface in the five-dimensional space $\{\mathbf{r},\chi,y\}$ it
remains of course spacelike). This effect is interpreted from the string
theory point of view as tilting the D-brane, but from the point of view of
3-dimensional field theory this multivaluedness  should be interpreted as a
signal that no well defined solution exists. It is worth noting that for BPS
solution such multivaluedness emerges for any value of the background field.

\begin{figure}[t]\centering
\includegraphics[width=11cm]{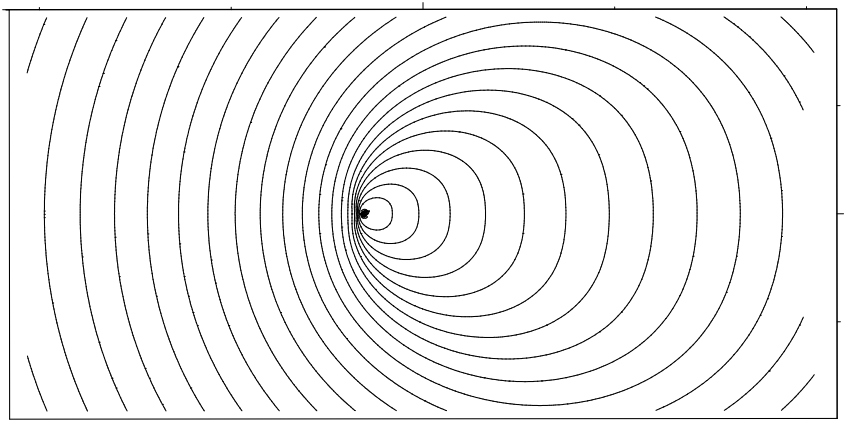}
\caption{Non-commutative $U(1)$ monopole: constant $|\mathbf{F}|$ curves}
\label{gal-fig:Fcurves}
\end{figure}

\begin{figure}[t]\centering
\includegraphics[width=11cm]{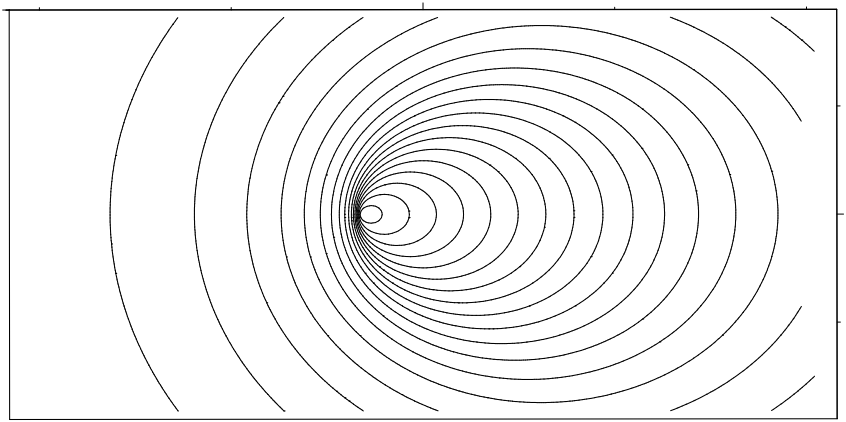}
\caption{Non-commutative $U(1)$ monopole: constant $y$ curves}
\label{gal-fig:Ycurves}
\end{figure}

In the figures \ref{gal-fig:Fcurves},\ref{gal-fig:Ycurves} the sections of
level surfaces of constant $y$ and constant $|\mathbf{B}|$ are shown. The
full solution is axially symmetric and is obtained by rotating the pictures
along the symmetry axis.

\section{Discussion}
We have discussed some new issues associated with the D-brane picture of
gauge theories. Apart from giving a nice geometric framework, D-branes
suggest a modification of the dynamics of the YM field suggesting the
Born-Infeld type lagrangian. This latter breaks the conformal invariance of
the YM equations removing the obstruction for existence of classical
glueballs in the SU(2) theory in four dimensions. Topological reason for
existence of such glueballs lies in the vacuum periodicity which holds
equally in the ordinary YM theory and in the NBI theory, with an important
difference that in the latter case the potential barriers between
neighboring vacua have finite heights. Classical NBI glueballs (more
precisely, half of them) are sphalerons mediating the topological
transitions. We have found that they exist both for the ordinary trace and
the symmetrized trace versions of the NBI theory with somewhat different
core structure. We have also shown that in the NBI theory with the triplet
Higgs one encounters, apart from the usual magnetic monopoles, the hybrid
solutions which can be regarded as sphaleronic excitations of monopoles.
Finally, adding the constant Kalb-Ramond field, one is able to account for
non-commutative monopoles. We presented a new nonperturbative axisymmetric
solution for the U(1) non-commutative monopole with Higgs.

\section*{Acknowledgements}

One of the authors (DG) is grateful to the organizers of the Workshop for
invitation and support and especially to Steven and Diana Duplij for a
stimulating atmosphere during this meeting. He is also grateful to LAPTH
(Annecy) for hospitality while the final version of this paper was
completed. This work was supported in part by the RFBR grant 00-02-16306.


\newpage

\end{document}